\newcommand{\kn}{ \mbox{KNbO$_3$} }
\newcommand{\bt}{ \mbox{BaTiO$_3$} }
\newcommand{\kt}{ \mbox{KTaO$_3$} }
\newcommand{\st}{ \mbox{SrTiO$_3$} }
\newcommand{\beq}{\begin{equation}}
\newcommand{\eeq}{\end{equation}}
\newcommand{\noi}{\noindent}

\documentstyle[preprint,aps,12pt]{revtex}
\begin{document}
\title{ Dynamic mechanisms of the structural phase transitions
 in KNbO$_{\bf 3}$: Molecular dynamics simulations }

\author{ M.Sepliarsky$^*$, M.G.Stachiotti$^*$, R.L.Migoni$^*$ and 
C.O.Rodriguez$^\dagger$  }
\address{$^*$ Instituto de F\'{\i}sica Rosario, Universidad Nacional de
              Rosario,\\
              27 de Febrero 210 Bis, 2000 Rosario, Argentina \\
         $^\dagger$ IFLYSIB, Grupo de F\'\i sica del S\'olido, C.C.565, 
              1900 La Plata, Argentina}

\maketitle

\begin{abstract}

The question on the dominant driving mechanism (displacive or 
order-disorder) at each structural phase transition of \kn\ is 
investigated by means of molecular dynamics simulations. 
To this purpose, we first develop a shell model by determining its potential
parameters in order to reproduce the ferroelectric instabilities obtained 
by first-principles total energy calculations.
The phase diagram as a function of temperature is obtained through
constant-pressure molecular dynamics simulations. 
The analysis of the dynamical structure factor and the microscopic dynamics   
of the particles in the different phases allows us to reveal the 
nature of the dynamics associated with each structural transition.
Correlations between local polarizations  
forming chain-like precursor clusters in the 
paraelectric phase are examined. 

\end{abstract}

\newpage

\section{Introduction}

Although the ferroelectric perovskite potassium niobate 
has been extensively studied by a variety of techniques,
the dynamical nature of its structural transitions remains 
an open question. This compound undergoes a sequence of three phase
transitions. With decreasing temperature it transforms from cubic 
paraelectric (C) to tetragonal ferroelectric (T) at 701 K, becomes 
orthorhombic (O) at 488 K, and finally rhombohedral (R) at 210 K.
The question, whether the dominant dynamical character at each transition
is displacive or order disorder, is of central importance, and numerous 
studies have been carried out in order to elucidate this point. 
 
A ferroelectric instability is characterised by a maximum of the potential
energy at the average positions of the ions in the high-temperature phase
with respect to collective displacements which lead to the low-temperature
phase. The average positions in the high-temperature phase correspond to
an energy barrier for the collective motion of the ions between equivalent 
average positions of the low-temperature phase. The dynamical character of
the transition is dominantly displacive or order-disorder depending on
whether such energy barrier is respectively low oder high compared with
the critical thermal energy $k_BT_c$. Within a purely displacive picture, 
there is a mode with the symmetry of the low-temperature phase which softens
with decreasing temperature and becomes unstable, i.e. its frequency vanishes
at the transition temperature.

Perovskites crystals have long been considered as displacive type 
ferroelectrics. The main evidence for this type of behavior has been the
existence of a $\Gamma$-TO soft mode which has been observed in many
perovskites$^{\cite{coc60,sco74}}$.

In the case of \kn, experimental evidences show that the sucessive C-T-O-R
phase transitions cannot be driven solely by optical-phonon softening, 
which was found to be incomplete by means of infrared 
spectroscopy$^{\cite{fon84}}$. Moreover, if hyper-Raman and infrared reflectivity spectra in 
the cubic and tetragonal phases of \kn\ are interpreted in terms of a soft 
mode, this has to be assigned such a high damping that it can hardly be 
distinguished from a relaxator$^{\cite{fon84,vog86}}$.
In fact, an unified interpretation of the experimental data has been
given by assuming the coexistence of a relaxational mode and a soft
phonon$^{\cite{fon84,fon88}}$. This picture has been supported by 
the observation of central components in light
spectra$^{\cite{vog86,fon88,sok88}}$ whose linewidths, lineshapes and symmetry
properties are consistent with an eight-site model of
the sucessive phase transitions.  
According to this order-disorder model, the potential energy surface has a
maximum for the cubic perovskite structure and eight degenerate minima
for the [111] displacements of the transition metal ion that correspond
to the low-temperature rhombohedral structure. In the cubic phase, the eight
sites are occupied with equal probability, and this symmetry is broken
as the temperature is lowered: four sites are occupied
in the tetragonal phase, two sites in the orthorhombic phase, and finally,
only one site is occupied in the rhombohedral structure.
In this way, the relaxation process which displays a critical slowing down
when T$_c$ is approached in the different phases is interpreted to be the
driving mechanism of the phase transitions.

Several features derived from other techniques support the above view.
Indeed, the most direct evidence that Nb atoms are displaced along the 
[111] directions, even in the paraelectric phase, comes from XAFS.
Pair distribution functions obtained by this technique in the 
different phases, show the presence of three short and three long Nb-O 
bonds which is consistent with a rhombohedral local structure and an
order-disorder mechanism for the the sucessive phase 
transitions$^{\cite{kim90,mat93,fre97}}$.  
However, in the orthorhombic phase of \kn\ a broad soft mode peak is clearly 
observed in  Raman experiments near the O-R transition, 
without any quasielastic component$^{\cite{fon88}}$, suggesting a more 
displacive-like dynamics for this transition. In addition, a more recent 
femtosecond time-resolved spectroscopy study for this phase 
rule out relaxational contributions of the same symmetry as the 
soft mode$^{\cite{dou94}}$. Also, other XAFS measurements at room temperature
revealed that the direction of the Nb displacement is as close to the polar
axis as the order of temperature atomic displacements$^{\cite{sku97}}$.

First principles calculations have contributed greatly to understand
the origins of the transitions in \kn$^{\cite{pos93,pos94,sin92,sin95,yu95}}$ 
and related perovskites$^{\cite{coh90,coh92,kin94,gho95,sin96}}$.
The local spin-density approximation (LSDA) has provided a useful framework
to elucidate important aspects of the underlying physics of these oxides, 
providing important quantitative information about 
electronic charge distributions, character of electronic bondings, crystal
structure, phonons, structural instabilities, etc., 
providing considerable insight into the nature of the soft-mode
total energy surface.

First-principles techniques are more powerful than any calculation 
based on empirical models. Nevertheless they are quite computer 
demanding and simulations of temperature driven structural transitions 
in perovskites are not computationally feasible at present.
A successful approach to study  finite temperature properties 
has been developed on the grounds of effective Hamiltonians whose 
parameters were fully determined from ab-initio calculations. 
This scheme has been applied to many ferroelectric 
perovskites$^{\cite{zho94,zho95,rabe97,kra98}}$.
However, the restricted dynamics considered, due to the lack of an atomistic
description, makes this approach inappropiate to investigate the anharmonic 
lattice dynamics of these compounds.

The goal of this work is to develop an atomistic model
which describes the dynamical properties and
phase transitions sequence of \kn\ in order to investigate 
which mechanism, displacive or order-disorder, dominates
in each structural phase transition.  
To this purpose, interatomic potentials are determined,
in the framework of a shell model, by comparing with ab-initio total energies 
calculations for different ferroelectric distortions.
The temperature behavior of the system and the dynamical nature of
its structural transitions are then investigated through 
molecular dynamics simulations.

\section{Model and computational details}

In spite of the above mencioned experimental evidence, the lattice dynamical
approach to ferroelectricity was primarily concerned with the microscopic
description of displacive ferroelectrics. In this way, 
an accurate description of the ferroeletric soft-mode regime in ABO$_3$ 
perovskites was given on the grounds of the 
non-linear oxygen polarizability (NOP) model$^{\cite{mig76}}$. 
This model emphasizes the large 
anisotropic polarization effects at the oxygens produced by variations of the 
O-B distance. Such effects are expected in view of the strong 
environment-dependent oxygen polarizability and its enhancement through 
hybridization between oxygen p and transition metal d orbitals.

In the NOP model, the $A$ and $B$ ions are considered 
isotropically polarizable.
On the other hand, an anisotropic core-shell interaction is considered for 
the O$^{-2}$ ion, reflecting its site symmetry. 
The above anisotropy is described by two different linear
core-shell coupling constants $K_{2}$: one in the directions of the
$A$ ions and another in the direction of the $B$ ion.
An additional fourth order contribution to the core-shell interaction
along the direction of the B ion is taken into account by a coupling
constant $K_{4}$.

The harmonic part of the model is unstable against the ferroelectric mode 
displacements in the cubic structure. This happens essentially because the 
strong Coulomb O-B attraction overcomes the repulsive forces that hold the B 
ion at the cubic cell center. Therefore the stability should be provided by the
fourth order term. This has been harmonically approximated by temperature 
averaging a pair of displacements, which is evaluated self-consistently. Hence 
the temperature dependence of the complete phonon dispersion can be calculated.
The effective harmonic term stabilizes the ferroelectric mode thus providing 
its temperature dependence. Applications to \kt$^{\cite{mig76,per89}}$, 
\st$^{\cite{mig76}}$, \kn$^{\cite{kug87}}$, \bt$^{\cite{kha89}}$, 
lead to results in 
excellent agreement with the experimental phonon data.

However, the self-consistent treatment of the quartic interaction in the 
NOP model washes out the potential maximum at the cubic structure in terms of
the soft mode coordinate. 
We have recently shown$^{\cite{sep95}}$ that an exact numerical
treatment of the fourth order interaction at oxygen
reproduces quite satisfactorily the ab-initio adiabatic potential
for the relevant ionic displacements which lead to the series of
ferroelectric transitions in \kn. 
A molecular dynamics simulation using this model allowed us
to confirm a crossover from a displacive to an order-disorder
character for the paraelectric-ferroelectric phase transition in
\kn$^{\cite{sep97}}$. However, due to the constant-volume treatment
performed, the simulation led to a direct transition from the
cubic to the lowest rhombohedral ferroelectric phase.

To obtain a model which describes the phase transition sequence of \kn,
it is necessary to allow for the homogenous strains involved in the phases;
a fact that was originally emphasized by Chaves et al.$^{\cite{cha76}}$.
This can be achieved by replacing harmonic force constants by interatomic
potentials. So, we first calculate potential parameters from the force 
constants by assuming nearest-neighbors pairwise $A-O$, $B-O$ and $O-O$
interactions. We choose to represent these by Buckingham potentials:
$V(r)~=~a~e^{(-\frac{r}{\rho})}~-~c~r^{-6}$. Actually, the Van der 
Waals term is included only for the $O-O$ interaction because it is 
attractive, as it turns out from the force constants. 
In a second  step, the model parameters are improved by fitting ab-initio 
total energy calculations for different ferroelectric distortions, with
and without lattice strain.

For the ab-initio total energy calculations we used the full-potential 
linear muffin-tin orbital (LMTO) method,
within the LDA and the Ceperly-Alder exchange-correlation potential.
Details of the method are given elsewhere$^{\cite{1msm,cor6}}$.
The calculations were performed employing
the same basis set used by Postnikov et al.$^{\cite{pos93}}$ and
the Brillouin zone integrations have been well converged in the
number of k-points.

For the investigation of the temperature driven structural transitions
we use constant-pressure molecular dynamics (MD) simulations. Shell model 
molecular dynamics, though it has a long history, is dificult to use due to 
the treatment of the adiabatic degree of freedoms, i.e. the shells.
Most workers have used a steepest descents method to relax the shell
positions iteratively to zero-force positions on each step of the molecular
dynamics. Although this procedure has been improved based on conjugate 
gradient relaxation of the shells$^{\cite{lin93}}$, in a typical 
simulation run an average of ten line searches are made within every time 
step of the simulation, reducing greatly the efficiency of the method in  
comparisson with rigid-ion model MD simulations.
An alternative approach has been introduced by Mitchell and 
Fichman$^{\cite{mit93}}$ in which the shells are given 
a small mass and their motion, like
those of the cores, is found by numerical integration of their equations of
motion. They showed that the results of this method are independent of the
shell mass, provided it is enough small,
and in agreement with those obtained using relaxation of massless
shells. Regarding the efficiency of the method, 
the shortcoming of this approach is that the time step of the 
simulation must be reduced in order to provide enough accuracy for the 
integration of the shell coordinates.

We applied the latter approach in the present MD simulation study, which is
carried out using the DL-POLY package$^{\cite{dlpoly}}$. 
The runs were performed employing a
Hoover constant-($\bar{\sigma}$,T) algorithm with external stress set to zero;
all cell lengths and cell angles were allowed to fluctuate.
Periodic boundary conditions over 4x4x4 primitive cells were considered; the 
basic molecular dynamics cell therefore contained 320 ions (plus 320 shells
which are additional degrees of freedom). The time step was
0.4 fs, which provided enough accuracy for the integration of the shell 
coordinates. The total time of each simulation, after 5 ps of thermalization, 
was 45 ps.

\section{Results}

\subsection{Potential determination and static properties of the model}

We have determined the model potential parameters in order to reproduce LMTO 
total energy calculations for different ferroelectric distortions, 
with and without lattice strain.
It is important to remark that one can not afford to reproduce very closely
the ab-initio energetics with such simple potentials as the ones employed in 
our model. In addition, the adjustment of the potentials   
is not a straightforward and easy 
procedure, because all pair potentials and core-shell coupling 
constants contribute to the total energy of a given distorted lattice 
structure.   

To evaluate the adiabatic energy surface for ferroelectric
distortions, we have calculated the potential energy of the model for
several atomic displacements.
To this purpose, the shell coordinates
(which represent the electronic degrees of freedom) are evaluated,
for a given core configuration, by solving the adiabatic condition.
Once the equilibrium solution for the shell coordinates is obtained
iteratively by a steepest descent procedure, the potential energy is 
computed.

As a result of the potential determination, we obtain the curves shown 
in Figure 1, where the results of the model are compared with LMTO 
calculations of the total energy as a function of the transition metal 
displacements, along [001], [011] and [111] directions.  
A reasonable agreement is achieved for the ferroelectric 
instabilities although the model leads to smaller
values for the off-center transition metal shifts.
In order to study the relevance of the lattice strain on the 
energetics, we performed displacements along the three mentioned 
directions for strained structures at the experimental values of 
$\frac{c}{a}$, $\frac{b}{a}$ and $\alpha$ (the rhombohedral strain angle).
These results are also shown in Figure 1.
The model reproduces satisfactorily the strong dependence of the energy
on the tetragonal and orthorhombic strains. We find  
a negligible effect on the total energies for 
the rhombohedral strain, so it is not plotted in the figure. 
Note that the relative values for the energy wells of the three strained 
structures is consistent with the experimentally observed phase 
transitions sequence. The model potential parameters 
obtained are listed in Table I. 

The bulk moduli, as in ab-initio calculations, can be obtained from the 
evaluation of the total energy as a function of the uniform volume expansion
for the cubic phase. The model calculations yield a lattice parameter of 
3.98 $\AA$ for the static perfect cubic structure. The bulk moduli evaluated 
at this equilibrium volume is 226 GPa, which agrees fairly well with the LDA
value of 208 GPa$^{\cite{pos93}}$.

\subsection{Phonon dispersion relations}

An interesting test of our modelling concerns the phonon dispersion 
relations. Recently, Yu and Krakauer$^{\cite{yu95}}$ have performed
first-principles phonon calculations for the ideal cubic
perovskite structure of \kn, using a linear response approach within
the framework of the LAPW method. This calculation reveals
structural instabilities with pronounced two-dimensional character
in the Brillouin zone, corresponding to chains of displaced Nb ions
oriented along the [001] directions. To check if our model is able
to reproduce such kind of instabilities, we compute the phonon
dispersion curves within the harmonic approximation for the Buckingham
potentials and retaining only the harmonic core-shell couplings at the
oxygen ions. The result is shown in Figure 2. Although a very good agreement 
is achieved for the stable modes compared with Ref.~\cite{yu95}, 
the imaginary phonon frequency
for the ferrolectric mode at $\Gamma$ turns out to be almost twice as
large in our model calculation as compared with the LAPW result.
This discrepancy could arise from the fact that the LMTO 
method produces an overestimation of the ferroelectric
instabilities in perovskites compared with LAPW results$^{\cite{sin96}}$.
Nevertheless, it is remarkable that the model reproduces the
wave vector dependence of the instabilities, as obtained by the 
ab-initio linear response approach (compare Fig.2 with Fig.1 
of Ref~\cite{yu95}), indicating chain-like instabilities in real space.
As highlighted by Yu and Krakauer$^{\cite{yu95}}$,
the finite thickness of the slab region of instability  
corresponds to a minimun correlation length of the displacement required
to observe an unstable phonon mode. From the model's phonon dispersion 
curves, the length of the shortest unstable chain can be estimated to
$\approx$ 4 $a$.

\subsection{Phase diagram}

We have used the above described model to perform constant-pressure MD 
simulations at several temperatures. In Fig. 3(a) we plot the order 
parameters (the three components of the mean polarization) as a function 
of temperature. The corresponding cell parameters are displayed in 
Fig. 3(b).

At high temperatures, the averaged polarizations
$p_x$, $p_y$ and $p_z$ are all very close to zero and the three lattice
constants have almost identical values. As the system is cooled down below 
725 K, $p_x$ acquires a value clearly different from zero, while  
$p_y \simeq p_z \simeq 0$, and the structure presents a considerable 
tetragonal strain (see Fig. 3(b)). This indicates the
transition from the paraelectric cubic to the ferroelectric tetragonal
phase. When the temperature is further reduced, the two lower ferroelectric
phases appear: the orthorhombic one below $\sim$ 475 K, with clearly finite
$p_x \simeq p_y $ and still $p_z \simeq 0$, and finally the rhombohedral
phase below $\sim$ 175 K, with approximately equal values of the three
polarization components. Although the model gives slighlty
large cell parameters and distortions compared with experimental
data$^{\cite{fon84}}$, the non-trivial phase transition sequence of \kn\ is
correctly reproduced. Regarding the transition temperatures, a good 
agreement with the experimental values is achieved. However, 
this can not be taken as a test on the quality of the model because
a larger size of 
the simulation supercell could increase the values of T$_c$. Thus a finite size 
scaling procedure would be necessary in order to determine the 
transition temperatures correctly. Such a treatment is beyond the scope 
of this work. 

Finally, we point out that the C-T and T-O transition temperatures obtained 
from our simulation are considerably larger than the ones obtained using the
effective Hamiltonian approach$^{\cite{kra98}}$. We believe there are two reasons
for such a discrepancy. The first one is the already mentioned  
overestimation of the ferroelectric instabilities produced by LMTO compared 
with LAPW. The second one is the thermal expansion which is taken into 
account in our approach. The thermal increase of volume stabilises the 
various phases over a wider range of temperatures.

\subsection{Dynamic character of the transitions}

To gain insight into the dynamic excitations relevant to
the ferroelectric phase transitions, we calculate the dynamical 
structure factor $S({\bf q},\omega)$ at the $\Gamma$ point.
After equilibration, $ S({\bf q}, \omega) $
is calculated with no thermostat and barostat, i.e. using conventional 
energy and volume conserving dynamics, as the space-time Fourier transform of 
the core-core displacement correlation function:
\begin{equation}
S({\bf Q},\omega)=\int_{-\infty}^{+\infty} dt e^{i \omega t}%
\sum_{l \kappa} \sum_{l^{\prime} \kappa^{\prime}}e^{ i{\bf Q}.%
\left( {\bf R}_\kappa^l-{\bf R}_{\kappa^{\prime}}^{l^{\prime}} \right) }%
\langle{\bf Q.u}_\kappa^l(t){\bf Q.u}_{\kappa^{\prime}}^{l^{\prime}}(0)\rangle
\label{strf}
\end{equation}
where {\bf Q} is an arbitrary wave vector, which can be decomposed in a
reciprocal lattice vector {\bf G} and a wave vector {\bf q} within the first 
Brillouin zone, {\bf Q} = {\bf G} + {\bf q}.
A gaussian smoothing procedure is applied before the time Fourier transform 
is performed.

On the left hand side of Figure 4 we show the low-frequency range of
$S({\bf q}=0,\omega)$ at 800 K (cubic phase), 525 K (tetragonal phase)
and 250 K (orthorhombic phase). In both the cubic and tetragonal phases a
quasielastic peak is observed, while no peak appears corresponding to the
ferroelectric (lowest TO) mode. The peaks
observed at $\approx$ 200 cm$^{-1}$ and $\approx$ 280 cm$^{-1}$ correspond
to the second lowest infrared active TO($\Gamma_{15}$) mode and the 
silent $\Gamma_{25}$ mode of the cubic phase, respectively.
The absense the soft-mode peak in our MD-spectra is consistent
with the spectroscopic observations mentioned in the 
Introduction$^{\cite{fon84,vog86,sok88}}$. Furthermore, in more recent 
inelastic neutron scattering measurements of cubic \kn,  the TO$_1$ optic
phonon peak was not detected for reduced wave vectors 
below 0.2$^{\cite{hol96}}$.
This experiment also showed the existence of an anomalously flat and 
low-energy TA mode, strongly coupled with the TO$_1$ mode at $q \sim 0.2$ and 
extending out to the Brillouin zone boundary. The analogous 
feature has been observed in \kt$^{\cite{per89}}$. 
Krakauer et al.$^{\cite{kra98}}$ 
obtained, through a MD simulation with an effective Hamiltonian, 
that the entire phonon branch softens in the cubic phase.  This observation 
of a single branch softening arises from the fact that the $\Gamma$-TO and
$X$-TA eigenvectors have very similar displacement patterns, aside from
the wave vector modulation, and they are 
described as a single effective phonon branch in their approach.   
Nevertheless, the dynamical behaviour of the effective degrees of freedom 
differs from the usual soft-mode picture of a displacive transition, and it 
is consistent with the previously obtained first-principles LAPW linear 
response results$^{\cite{yu95}}$.

To clarify the nature of the microscopic dynamics leading to the above 
remarqued features of $S({\bf q}=0,\omega)$, we show on the right hand side 
of Figure 4 the time evolution of a single cell polarization component (a
vanishing one in average) at the same temperatures. We observe for both the
cubic and tetragonal phases that fast oscillations
around finite polarization values coexist with much slower polarization
reversals. So, the dynamics posseses two components with different time 
scales. While one component is associated with quasi-harmonic oscillations
around an off-center position, the other refers to a relaxational
motion between equilibrium sites. It is clear that the dynamics associated 
with the appearence of the quasi-elastic component in the cubic phase is a
relaxational motion of local polarizations between the eight energy minima 
along [111] directions. In the tetragonal phase, one component ($p_x$) 
remains oscillating around a non-zero value while the other two show a
relaxational dynamics between four energy minima. 
This dynamical behavior is in agreement with the picture provided by
the eight site model and fully agrees with the results of a previous MD 
simulation study using an interacting 
polarizable ions model$^{\cite{edw89}}$.  
So, we can conclude that a relaxational slowing-down
proccess is therefore mainly responsible for the C-T and T-O phase
transitions in \kn.

While the ferroelectric transition (C-T) is generally discussed in terms of 
order-disorder models with relaxational-type dynamics, aharmonic damping 
effects are less severe at room temperature and below, and a simple soft-mode
description appears to be 
adequate for the O-R phase transition$^{\cite{fon88,dou94}}$. 
The dynamical structure factor for the orthorhombic phase
of \kn, plotted in Figure 4, shows a broad phonon peak centered at 
$\approx$ 100 cm$^{-1}$, which is assigned to the ferroelectric mode, 
while no central component appears. The absence of a quasi-elastic peak 
would indicate that a different microscopic dynamics govern the O-R phase
transition. In fact, a more oscillatory dynamics is observed for the 
non-polar coordinate in the O phase (see the righ hand side of Figure 4). 
To rule out the effects introduced by the small overstimation of the lattice
strain obtained through our modelling approach, we have performed the same
simulation study but using the experimental values of the lattice constants. 
In this way the picture obtained is qualitatively similar to the one reported, 
showing an oscillatory dynamics without the presence of a quasi-elastic 
peak in the spectrum. This fact clearly indicates a more displacivelike
dynamics for the O-R transition.

\subsection{Dynamic correlations in the cubic phase}

How the polarization of a single unit cell correlates with its 
neighbors, in the proximity of the ferroelectric phase transition, is a 
subject of central importance. The discovery of 2D X-ray diffuse intensity 
patterns and the systematic disappearence of them during the transitions
have led Comes et al.$^{\cite{com70}}$ to suggest the formation of
static linear chains in real space, where the Nb atoms are displaced 
along [111] directions within each cell. Latter on, from neutron scattering
experiments on KTaO$_3$, an alternative explanation was provided
based on a flat dispersion of the TA phonon branch along [100] sheets of the
Brillouin zone$^{\cite{com72}}$. These results suggested that the linear 
correlation of atomic displacements along the [100] directions are dynamic, 
in agreement with a dynamical model introduced by H\"{u}ller$^{\cite{hul69}}$
a couple of years before. More recent high-resolution diffuse x-ray
scattering measurements using synchrotron radiation have supported the 
dynamical model of correlations$^{\cite{hol95}}$.

The existence of correlated atomic displacements forming chains have been
also supported for the already mentioned ab-initio linear response 
calculation of Yu and Krakauer$^{\cite{yu95}}$, which showed the existence of 
Brillouin zone planar instabilities in the cubic phase of \kn. 
Recent MD simulations, using the effective Hamiltonian approach, 
have revealed preformed dynamic chain-like structures that are related to 
the softening of a phonon branch over large 
regions of the Brillouin zone$^{\cite{kra98}}$.  

As it was showed in Section III-B, our model reproduces the wave vector
dependence of the instabilities indicating, as obtained by the linear response
approach, chain-like instabilities in real space. To further confirm this point
we have calculated correlation functions for the three components of the local
polarization as a function of the cell-cell distance d along a z-axis chain.
This was done by defining the following correlation functions:

\begin{equation}
{\it P}_{xx}(d) = < {\rm P}^{i}_{x}(t) {\rm P}^{i+d}_{x}(t) >  
\end{equation}
\begin{equation}
{\it P}_{yy}(d) = < {\rm P}^{i}_{y}(t) {\rm P}^{i+d}_{y}(t) >  
\end{equation}
\begin{equation}
{\it P}_{zz}(d) = < {\rm P}^{i}_{z}(t) {\rm P}^{i+d}_{z}(t) >  
\end{equation}

\noi where ${\rm P}^{i}_{\alpha}(t)$ and ${\rm P}^{i+d}_{\alpha}(t)$ are 
the $\alpha$ component of the instantaneous local polarization of cell $i$ 
and $i+d$, respectively, with the condition that both cells belong to the 
same z-axis chain. The correlation function ${\it P}_{\alpha \alpha}(d)$ 
is defined as the space-time average of the product of those equal time 
local polarizations. These functions were calculated after equilibration,
with no thermostat and barostat, using periodic boundary conditions over 
8x8x8 primitive cells. The results at T=800~K are shown in Figure 5. While 
${\it P}_{xx}(d)$ and  ${\it P}_{yy}(d)$ decrease strongly when the cell-cell 
distance (d) increases,  ${\it P}_{zz}(d)$ shows a slow decrease. This 
clearly indicates that local polarization components parallel to a given 
chain are highly correlated along that chain, while the perpendicular 
components are fully uncorrelated. An estimation of the correlation lenght 
can be done by fitting the points of Figure 5 to the function 
${\it P}_{\alpha \alpha}(d) = e^{-\frac{d}{\xi_{\alpha \alpha}}}$. However,
due to the small size of the simulation supercell, the correlations of 
relatively more distant cells are affected by the periodic boundary 
conditions, particularly for longitudinally correlated cells. Therefore we
use only the point correponding to the smallest d for the determination of
$\xi_{zz}$. This yields $\xi_{zz} \approx 6 a$ and 
$ \xi_{xx} \approx \xi_{yy} \approx 0.5 a$, and the corresponding exponentials
are shown in Figure 5.

To visualize chain-like correlations in real space, we show in
Figure 6 snapshots of instantaneous local configurations, in an arbitrary 
slice through the simulation supercell, at three different times. The slice 
was choosen with the z-axis as the vertical one. Three symbols distinguish the
values of the z-coordinate of the local polarizations: $\uparrow$ when 
$P^i_z > 15 \frac{\mu C}{cm^2}$, $\downarrow$ when 
$P^i_z < -15 \frac{\mu C}{cm^2}$ and $\circ$ when 
$| P^i_z | < 15 \frac{\mu C}{cm^2}$ (thus discarding as non z-polarized
the cases where $| P^i_z |$ is small). 
This plot allows to visualize clearly
chain-like correlations of $P^i_z$ through
the appearance of finite lenght chains polarized up or down along the z-axis.
It is dificult to estimate quantitatively the lenght of the chains due to 
their dynamical evolution, but some of them extend over the whole 
supercell width. It is interesting to point out that cells belonging to 
a given chain seem to correlate their motion, producing coherent 
polarization reversals.

To further confirm this point we have plotted in Figure 7 the time evolution 
of $P^i_z$ and $P^i_x$ of four consecutive cells which belong to a given 
z-axis chain. Chain-like correlations are evident due to the correlated motion
(polarization reversals) of $P^i_z$ and the fully uncorrelated dynamics of
$P^i_x$.

\section{Conclusion}

We have developed an atomistic model for \kn\ which describes its structural
instabilities in good agreement with LMTO total energy calculations. A further
molecular dynamics simulation allowed us to evaluate the phase diagram,
reproducing correctly the non-trivial phase transition sequence of \kn. 

Regarding the dynamical mechanisms of the phase transitions, 
were able to identify the dominant character of each one.
We find that the 
C-T and T-O transitions have a predominant order-disorder character, 
signed by the presence of a central peak in the dynamic response spectra. 
This excitation appears because of the slow dynamics associated with a 
relaxational motion of local polarizations, which correlate 
within chain-like precursor domains in the paraelectric phase. 
On the other hand, the O-R
transition is more displacivelike, showing a more oscillatory dynamics of 
the non-polar coordinate.

\vspace{2.cm}
\noi { Acknowledgments} \\
We thank A. Dobry for helpful discussions.
This work was supported by the
Consejo Nacional de Investigaciones Cient\'{\i}ficas y T\'ecnicas
de la Rep\'ublica Argentina. M.S. acknowledges support from
Consejo de Investigaciones de la Universidad Nacional de Rosario.

\figure{Figure 1: Total energy versus Nb displacements, obtained at the
experimental lattice constant, with and without tetragonal and
orthorhombic strains. The energies are referred to the cubic structure.
The unstrained crystal results are represented with full lines for the
model and circle points for the LMTO calculations. The results for the
strained lattices are shown by dotted lines for the model and
triangule points for the LMTO.      }

\figure{Figure 2: Calculated phonon dispersions in the
cubic structure at the experimental lattice constant. Imaginary phonon
frequencies are represented as negative values.  }

\figure{Figure 3:  Phase diagram of KNbO$_3$: a) the three components of the 
average polarization as a function of temperature. b) the corresponding cell
parameters.}

\figure{Figure 4: Dynamical structure factor $S({\bf q}=0, \omega)$ and time 
evolution of the z-component polarization in a single cell at several 
temperatures. }

\figure{Figure 5: Correlation functions for the three components of the local
polarization as a function of the cell-cell distance (d) along a z-axis chain
in the cubic phase of \kn\ (T~=~800~K).

\figure{Figure 6: Snapshots of instantaneous local configurations, in an
arbitrary slice through the simulation supercell, at three different times. 
Three symbols distinguish the values of the z-coordinate of the local
polarizations: $\uparrow$ when $P^i_z > 15 \frac{\mu C}{cm^2}$ , 
$\downarrow$ when $P^i_z < -15 \frac{\mu C}{cm^2}$ and 
$\circ$ when $| P^i_z | < 15 \frac{\mu C}{cm^2}$. }

\figure{Figure 7: Time evolution of the z-component (a) and x-component (b)
of local polarizations for four consecutive cells which belong to a given 
z-axis chain.}

\newpage

\begin{table} 
\caption{Potential parameters of the shell model. a,$\rho$,c: Buckingham
parameters; Z,Y: ionic and shell charge; $K_2$,$K_4$: on site core-shell 
force constants. The symbols $\parallel$ and $\perp$ refer to parallel and
perpendicular directions to the Nb-O bond, respectively. } 
\vspace*{0.3cm}         
\label{table1}
\begin{tabular}{|cccc|ccccc|}
Interaction & a (eV) & $\rho { (A)}$ & $c{ (eVA}^{-6}$) & Ion & Z($%
\left| e\right| )$ & Y($\left| e\right| )$ & $K_2({eVA}^{-2})$ & $K_4%
({eVA}^{-4})$ \\ \hline
\multicolumn{1}{|l}{${K - O}$} & \multicolumn{1}{r}{$124872.44$} & 
\multicolumn{1}{l}{0.19499} & \multicolumn{1}{c|}{0.0} & \multicolumn{1}{l}{
K} & \multicolumn{1}{r}{0.82} & \multicolumn{1}{r}{-0.42} & 
\multicolumn{1}{l}{$225$} & \multicolumn{1}{l|}{} \\ 
\multicolumn{1}{|l}{${Nb-O}$} & \multicolumn{1}{r}{$1036.63$} & 
\multicolumn{1}{l}{0.38997} & \multicolumn{1}{c|}{0.0} & \multicolumn{1}{l}{
Nb} & \multicolumn{1}{r}{4.84} & \multicolumn{1}{r}{7.82} & 
\multicolumn{1}{l}{288} & \multicolumn{1}{l|}{} \\ 
\multicolumn{1}{|l}{${O - O}$} & \multicolumn{1}{r}{$3597.22$} & 
\multicolumn{1}{l}{$0.34659$} & \multicolumn{1}{c|}{800.0} & 
\multicolumn{1}{l}{O} & \multicolumn{1}{r}{-1.88} & \multicolumn{1}{r}{-3.01
} & \multicolumn{1}{l}{$66.00\parallel $} & \multicolumn{1}{l|}{$%
300\parallel $} \\ 
\multicolumn{1}{|l}{} & \multicolumn{1}{l}{} & \multicolumn{1}{l}{} & 
\multicolumn{1}{l|}{} & \multicolumn{1}{l}{} & \multicolumn{1}{l}{} & 
\multicolumn{1}{l}{} & \multicolumn{1}{l}{$92.25\perp $} & 
\multicolumn{1}{l|}{} \\ 
\end{tabular}
\end{table}

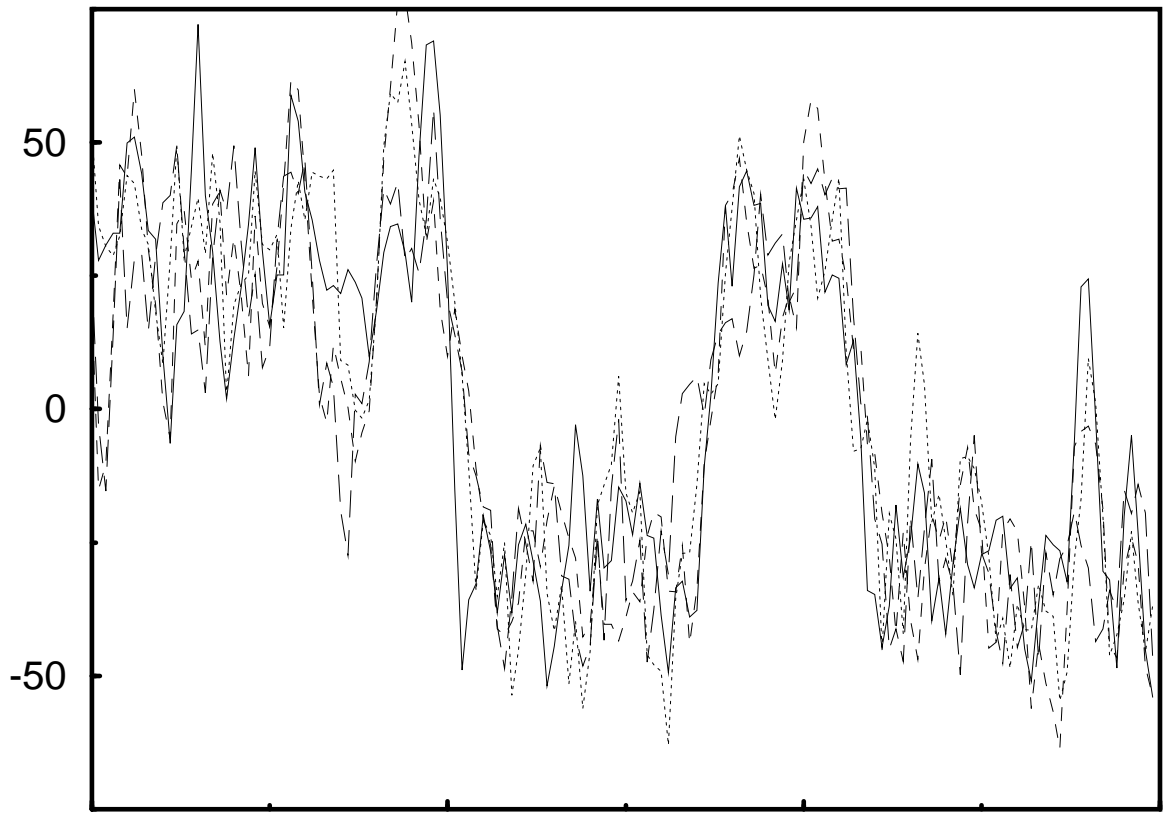
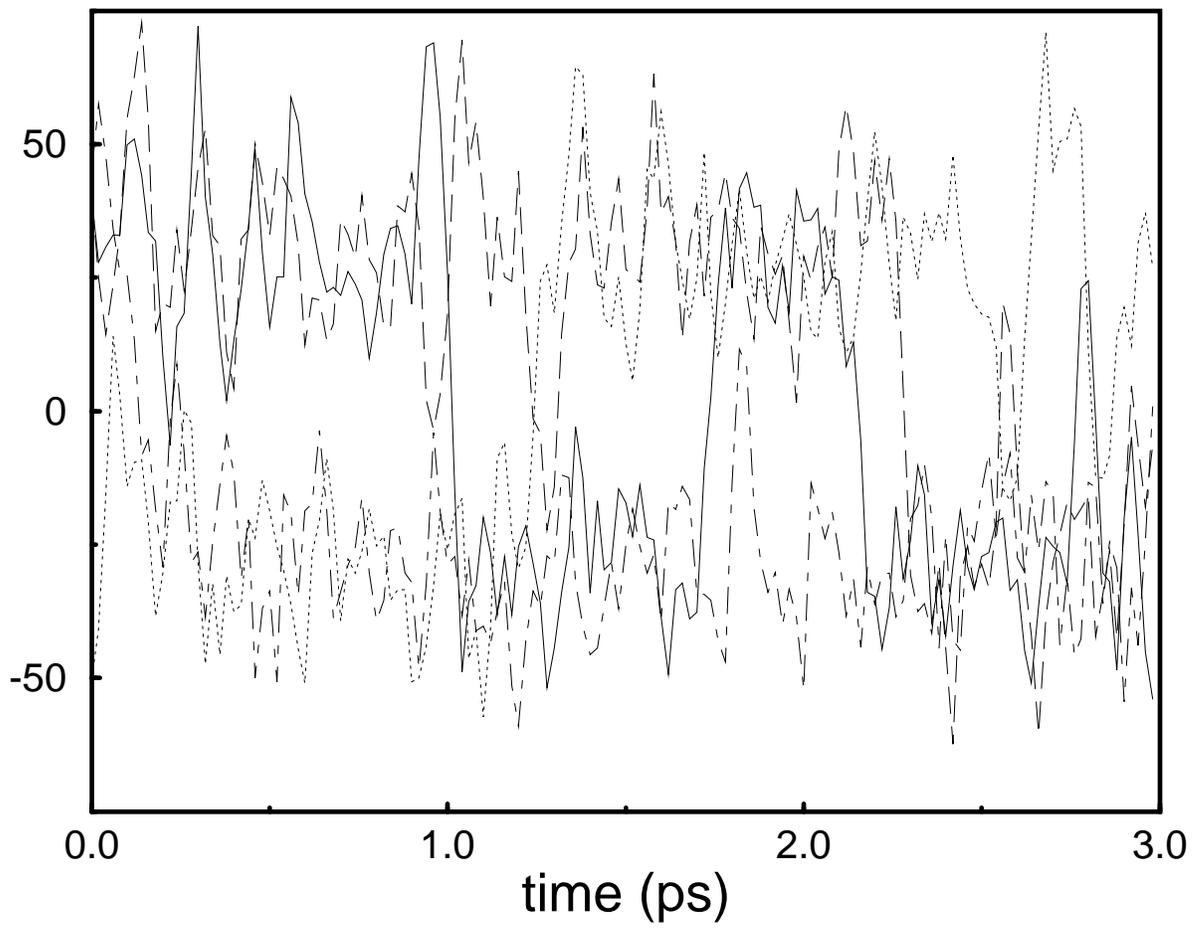
\end{document}